\documentclass[11pt]{article}

\usepackage[utf8]{inputenc}
\usepackage[T1]{fontenc}
\usepackage[english]{babel}
\usepackage[caption=false]{subfig}
\usepackage{amsmath}
\usepackage{amssymb}
\usepackage{graphicx}
\usepackage{rotating} 
\usepackage{caption}
\usepackage{float}
\usepackage{geometry}
\usepackage{setspace}
\usepackage{xkeyval}
\usepackage{booktabs}
\usepackage{color}
\usepackage{epstopdf}
\usepackage{authblk}
\usepackage[super,sort&compress,comma]{natbib} 

\usepackage{geometry}
\title{\LARGE{\vspace{-2.0cm}\textbf{Rubidium intercalation in epitaxial monolayer graphene}}}

\author[a,$\star$]{Letizia Ferbel}
\author[a]{Stefano VeronSI}
\author[b]{Tevfik Onur Mentes}
\author[c]{Lars Bu{\ss}}
\author[d]{Antonio Rossi}
\author[d]{Neeraj Mishra}
\author[d]{Camilla Coletti}
\author[c]{Jan Ingo Flege}
\author[b]{Andrea Locatelli}
\author[a]{Stefan Heun}

\affil[a]{NEST, Istituto Nanoscienze-CNR and Scuola Normale Superiore, Piazza S. Silvestro 12, 56127 Pisa, Italy}
\affil[b]{Elettra-Sincrotrone Trieste S.C.p.A., Strada Statale 14, km 163.5, I-34149 Basovizza, Trieste, Italy}
\affil[c]{Applied Physics and Semiconductor Spectroscopy, Brandenburg University of Technology Cottbus-Senftenberg, 03046, Cottbus, Germany}
\affil[d]{Center for Nanotechnology Innovation@NEST, Istituto Italiano di Tecnologia, Piazza S. Silvestro 12, 56127 Pisa, Italy}

\affil[$\star$]{\textnormal{letizia.ferbel@sns.it}}

\date{}                     
\begin{document}
\maketitle

\noindent{\textbf{Abstract. }Alkali metal intercalation of graphene layers has been of particular interest due to potential applications in electronics, energy storage, and catalysis. Rubidium (Rb) is one of the largest alkali metals and the one less investigated as intercalant. Here, we report a systematic investigation, with a multi-technique approach, of the phase formation of Rb under epitaxial monolayer graphene on SiC(0001). We explore a wide phase space with two control parameters: the Rb density (i.e., deposition time) and sample temperature (i.e., room- and low-temperature). We reveal the emergence of $(2 \times 2)$ and $(\sqrt{3} \times \sqrt{3})$R30° structures formed by a single alkali metal layer intercalated between monolayer graphene and the interfacial C-rich reconstructed surface, also known as buffer layer. Rb intercalation also results in a strong n-type doping of the graphene layer. Progressively annealing to high temperatures, we first reveal diffusion of Rb atoms which results in the enlargement of intercalated areas. As desorption sets in, intercalated regions progressively shrink and fragment. Eventually, at approximately 600~°C the initial surface is retrieved, indicating the reversibility of the intercalation process.} \\

\section{Introduction}

Alkali metals (AMs) are relevant in many fields due to their low electronegativity, high reactivity, and catalytic properties. 
The research on AMs intercalation in between graphene layers started in the early 1920s with graphite intercalation compounds (GICs).\cite{Fredenhagen1926} 
The insertion of AMs between graphene layers presents promising opportunities for use in electronics, energy storage, and catalysis.\cite{Inagaki1989, Xu2017, Bisio2024} However, understanding the details of AM intercalation at specific graphene interfaces remains a priority.

Lithium (Li), sodium (Na), and potassium (K) are widely studied for their application in rechargeable ion batteries and electrical conductors.\cite{Xu2017,Li2019} Intercalation of potassium in graphene has also sparked interest for its superconductivity at relatively high temperatures~\cite{Xue2012} and the extended van Hove singularities in the graphene band structure around the M point,\cite{McChesney2010} while cSIum (Cs) intercalated layers were demonstrated to realize a two-dimensional Fermi gas.\cite{Hell2020} Interestingly, Rb-intercalated bilayer graphene was shown to exhibit a metallic interlayer state,\cite{Kleeman2013,Kleeman2014,Kaneko2017} which is associated with superconductivity in intercalated graphene compounds.\cite{csanyi2005, Durajski2019} Furthermore, intercalated rubidium (Rb) has come under focus in the advancing field of ultra-cold-atoms dispensers.\cite{Kohn2020} Rb also appears to be very promising and worth investigating in the field of energy storage.\cite{Baichuan2023}

\begin{figure}[h!]
	\centering
	\includegraphics[width=0.5\linewidth]{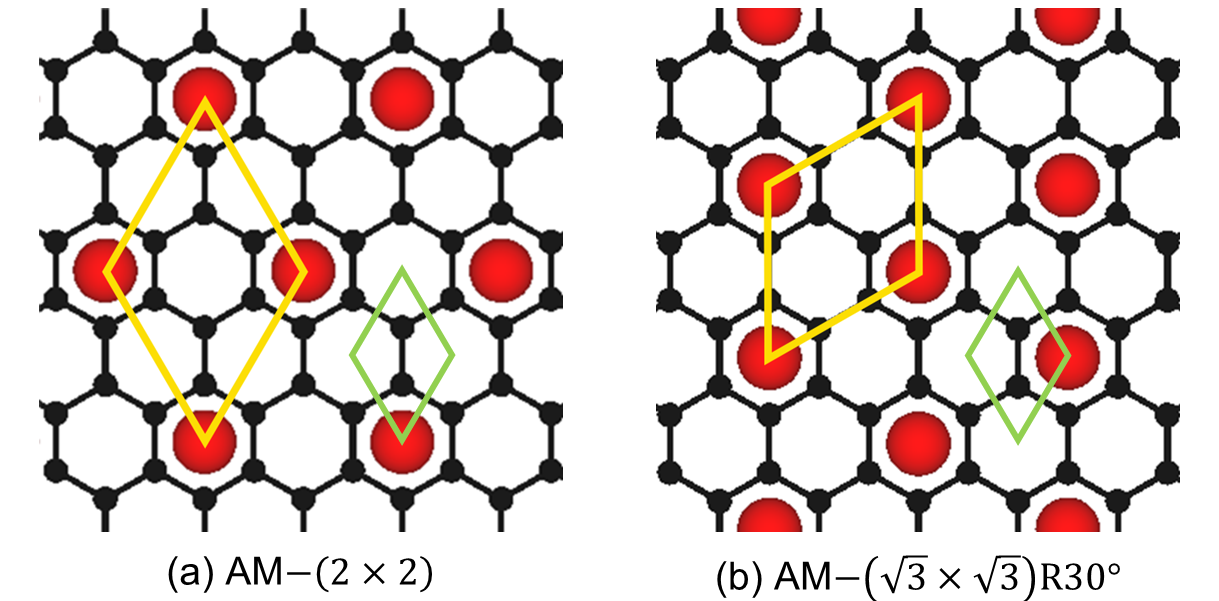}%
	\caption{Schematic representation, in top view, of the AM distribution (a) in the $(2 \times 2)$ superstructure, and (b) in the $(\sqrt{3} \times \sqrt{3})$R30° superstructure. Black circles are C atoms, while red circles are AM atoms. The graphene unit cell is highlighted by a green rhombus while the unit cells of the superstructures are highlighted in yellow.}
	\label{Fig1}
\end{figure}

A clear understanding of the atomic arrangement of AM atoms between graphene layers is crucial for optimizing such graphene-based devices, as this influences the electronic structure and thus the final properties of the material. Highly ordered structures of AM intercalants have been reported to appear in carbon-layered structures ranging from bilayer graphene up to bulk graphite. In these systems, K, Rb, and Cs atoms form a $(2 \times 2)$ superstructure with respect to the graphene lattice~\cite{Caragiu2005, Algdal2007, Kleeman2013, Petrovic2013, Li2019, Huempfner2022} (sketched in Fig.~\ref{Fig1}(a)), while the intercalation of Li leads to a more densely packed superstructure of $(\sqrt{3} \times \sqrt{3})$R30° periodicity with respect to the graphene lattice~\cite{Fiori2017, Li2019} (sketched in Fig.~\ref{Fig1}(b)). 
However, the ordered Rb intercalation of graphene monolayers has not yet been reported. So far, only the formation of an ordered Rb$(2 \times 2)$ \textit{over}layer on quasi-free-standing monolayer graphene on SiC(0001) was reported.\cite{Shin2020}

Here we reveal, with microscopic detail, the processes of intercalation and de-intercalation of Rb underneath monolayer graphene on SiC(0001). We study the Rb intercalation morphologies and de-intercalation dynamics by using in-situ conventional and micro-spot low-energy electron diffraction (LEED and $\mu$-LEED), scanning tunneling microscopy (STM), and low-energy electron microscopy (LEEM) in conjunction with density functional theory (DFT). Our results shed light on the Rb intercalation mechanism below graphene and demonstrate that 2D-ordering of intercalated Rb atoms can be achieved even in monolayer graphene.  

\section{Experimental}

Graphene was epitaxially grown on nominally on-axis n-type 6H--SiC(0001) wafers. The graphene growth was obtained in a BM-Aixtron reactor via silicon sublimation at temperatures of 1250--1300~°C in argon atmosphere.\cite{Emtsev2009, Rossi2018} Here we use surfaces consisting of a mixture of buffer-layer and monolayer graphene regions. The graphene quality, uniformity, and composition were first assessed in air by atomic force microscopy and Raman spectroscopy. Once in the ultra-high-vacuum chamber (UHV, base pressure $<1\times 10^{-10}$~mbar), the as-grown graphene samples were degassed at 600~°C to eliminate adsorbates. Subsequently, the graphene quality was further verified in UHV via LEED and STM, and in some cases also by LEEM. 

Rb was evaporated from a commercial dispenser (SAES Getters Inc.) onto the graphene surface held either at room-temperature (RT) or at low-temperature (LT, 100-140~K). The Rb deposition time accounts for the time the sample is directly facing the Rb evaporator after reaching the dSIred deposition conditions (evaporator current and flux were kept constant during deposition). Due to the different geometry of the evaporator set-up, the Rb yield in the LEEM experiment was $\sim 60$ times lower than the Rb yield in the STM experiment. Rb diffusion and de-intercalation were achieved by annealing the samples for 10~min at temperatures in the range of 50--800~°C in the case of RT-Rb deposition and in the range of 160--300~K in the case of LT-Rb deposition.

STM data were acquired with a VT-RHK-STM operating in constant current mode, at RT, and under UHV conditions. STM images were processed with the Gwyddion software package.\cite{Necas2012} The STM preparation chamber is equipped with a commercial LEED OCI BDL-600IR (spot size $\sim$500$~\mu$m). 

LEEM (resolution $\sim10$~nm) and $\mu$-LEED (spot size $\sim$1$~\mu$m) measurements were performed in a SPELEEM III (Elmitec GmbH) setup at the Nanospectroscopy beamline (Elettra, Trieste).\cite{Locatelli2006, Mentes2014}

DFT calculations were performed using the projector augmented wave method as implemented in GPAW~\cite{mortensen2024} using the Perdew-Burke-Ernzerhof (PBE)~\cite{perdew1996a} parameterization of the Generalized Gradient Approximation (GGA). Dispersion corrections were included using the Grimme D4 model (DFT-D4). \cite{caldeweyher2017,caldeweyher2019,caldeweyher2020} A $7 \times 7 \times 1$ Monkhorst-Pack grid was used for integration of the 3D Brillouin zone and a cut-off energy of 400~eV was used for the plane wave basis set. A detailed discussion of the calculations is available in the SI.

\section{Results}

\begin{figure}[b!]
	\centering
	\includegraphics[width=\linewidth]{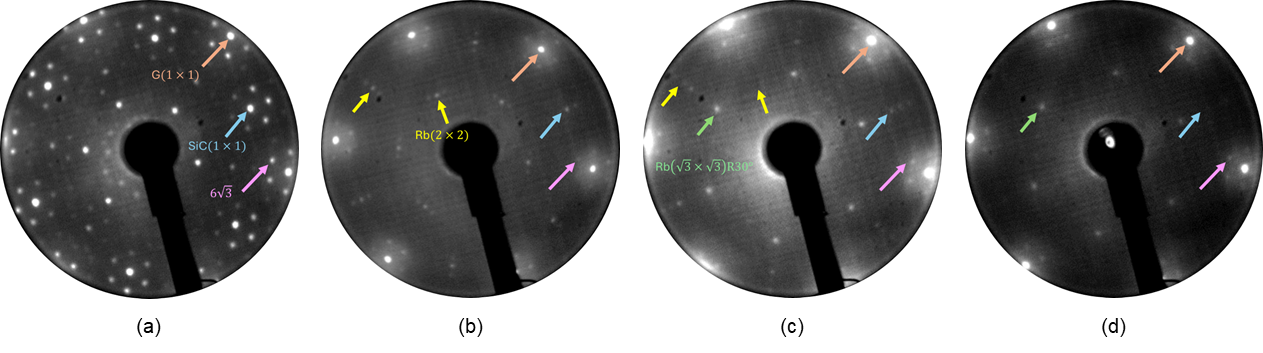}%
	\caption{Evolution of the LEED pattern upon Rb deposition on epitaxial graphene on SiC(0001) at room temperature. LEED pattern obtained from (a) the pristine graphene sample, and after depositing Rb for (b) 1~min, (c) 2~min, and (d) 3~min. The graphene, SiC, and $6\sqrt{3}$ structures are indicated by peach, blue, and pink arrows, respectively. The Rb$(2\times2)$ and Rb$(\sqrt{3}\times \sqrt{3})$R30° structures are highlighted by yellow and green arrows, respectively. Beam energy: (a) 60~eV, (b) 75~eV, (c) 60~eV, and (d) 65~eV.}
	\label{Fig2}
\end{figure}  

Figure~\ref{Fig2} shows the coverage-dependent evolution of the graphene LEED pattern as obtained after Rb deposition at room temperature. 
The diffraction pattern of the pristine sample (shown in Fig.~\ref{Fig2}(a)) shows the $(1\times1)$ graphene and $(1\times1)$ SiC substrate patterns as well as the characteristic $(6\sqrt{3}\times 6\sqrt{3})$R30° moirè reconstruction (that will be referred to as $6\sqrt{3}$ in the following) originating from the interaction and partial bonding of the buffer layer with the SiC substrate.~\cite{Riedl2010}
After Rb deposition (Fig.~\ref{Fig2}(b)--(d)), the diffraction spots of the pristine sample change their relative intensities. The SiC and $6\sqrt{3}$ diffraction intensities weaken, but these spots do not disappear. On the other hand, the intensity of the graphene spots does not change and remains the brightest, suggesting the presence of Rb atoms below the graphene layer rather than adsorbed. The presence of the $6\sqrt{3}$ pattern excludes the presence of Rb below the buffer layer. Consequently, this data suggests the presence of a Rb interlayer between the buffer layer and the graphene, which attenuates the diffraction intensity of the layers below it.
After Rb deposition for 1~min (shown in Fig.~\ref{Fig2}(b)), in the LEED pattern an additional set of diffraction spots appears. These correspond to an ordered Rb superstructure with $(2\times2)$ periodicity with respect to graphene. 
With further Rb deposition (2~min, shown in Fig.~\ref{Fig2}(c)), a second set of additional diffraction spots develops. These are rotated by 30° compared to the graphene lattice and identified as a $(\sqrt{3} \times \sqrt{3})$R30° superstructure, corresponding to an ordered Rb structure with higher density. 
The diffraction patterns of these two Rb ordered superstructures coexist up to a saturation coverage, above which only the $(\sqrt{3} \times \sqrt{3})$R30° structure remains (shown in Fig.~\ref{Fig2}(d)). 
These changes in the diffraction pattern indicate the formation of a well-ordered alkali metal layer sandwiched between the buffer layer and monolayer graphene whose structure evolves from a $(2\times2)$ to a $(\sqrt{3} \times \sqrt{3})$R30° with increasing Rb coverage at RT.

At RT, the $(2\times2)$ superstructure is not stable. After about 20~min from the Rb deposition, all features in the LEED pattern originating from the $(2\times2)$ structure disappear completely. On the other hand, the denser $(\sqrt{3} \times \sqrt{3})$R30° superstructure remains stable for at least several months under UHV conditions at RT.

Low-temperature Rb deposition followed by annealing cycles up to room-temperature also results in a continuous evolution of the Rb interlayer from a $(2\times2)$ superstructure to a $(\sqrt{3} \times \sqrt{3})$R30° superstructure (as reported by the diffraction analysis shown in Figs.~S1 and S2 of the SI). By depositing Rb at 100-140~K, a low intensity and diffuse $(2\times2)$ pattern develops. By progressively heating the sample, the $(2\times2)$ reflections get sharper while the diffuse background decreases, indicating an increased ordering of the intercalated phase. At 190~K the $(\sqrt{3} \times \sqrt{3})$R30° pattern appears, in coexistence with the $(2\times2)$ reconstruction. The $(2\times2)$ superstructure vanishes close to RT, while the $(\sqrt{3} \times \sqrt{3})$R30° superstructure is visible in a wide temperature range above 190~K, including room-temperature. This evolution of the LEED pattern suggests that initially a large amount of deposited Rb is adsorbed on the surface in a disordered manner. Diffusion is induced by annealing which allows the phase transition between the $(2\times2)$ and $(\sqrt{3} \times \sqrt{3})$R30° superstructures.

\begin{figure}[hbt!]
	\centering
	\includegraphics[width=0.5\linewidth]{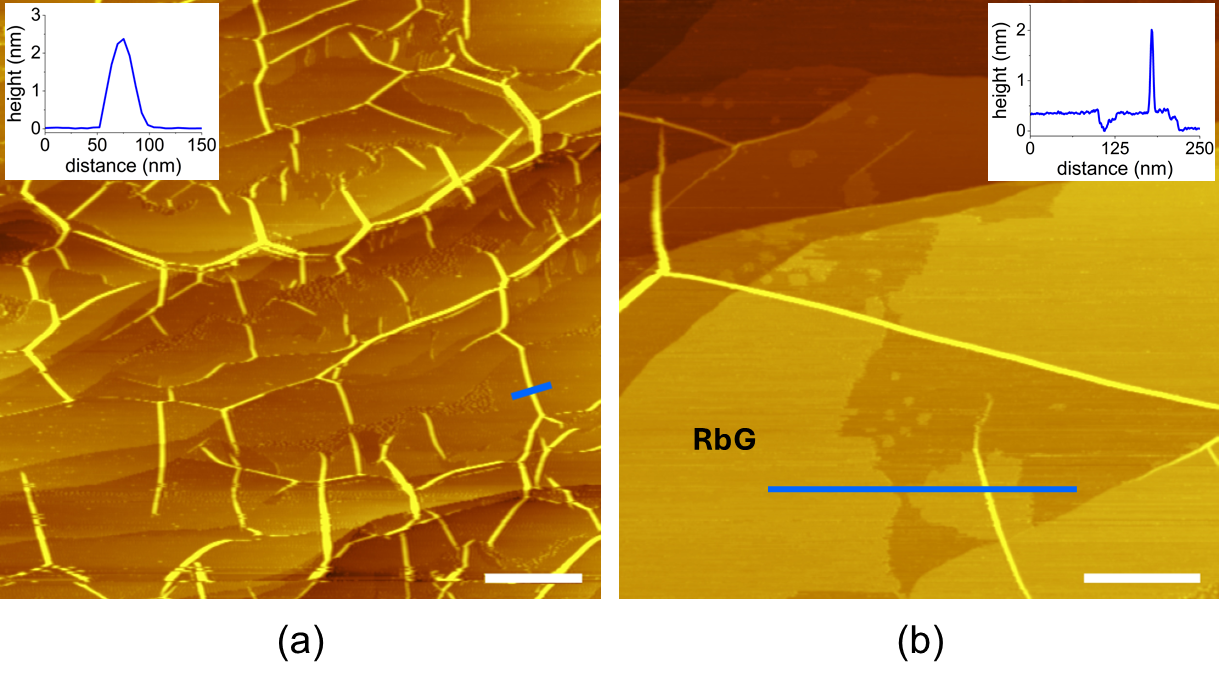}%
	\caption{(a) Overview STM scan of the graphene surface obtained after Rb deposition for 3~min at RT showing a wrinkles network. Inset to (a): Cross-section taken across a wrinkle, along the blue line in (a). (b) Close-up view of monolayer graphene showing the wrinkles network and Rb intercalated regions, labeled as RbG. Inset to (b): Cross-section taken across RbG regions, along the blue line in (b). Scan size: (a) $(3~\mu\text{m}\times3~\mu\text{m})$ and (b) $(500\text{~nm}\times500\text{~nm})$. Scale bar: (a) 500~nm and (b) 100~nm.}
	\label{Fig3}
\end{figure}

Further information is obtained from STM imaging. A few key modifications of the graphene surface due to Rb deposition can be identified. Figure~\ref{Fig3} reports the typical large-scale surface morphology obtained after Rb deposition at RT. As shown in Fig.~\ref{Fig3}(a), the step terrace morphology of the pristine graphene/SiC sample can still be easily recognized, in addition to a wrinkle network that appears all over the monolayer graphene surface, but that avoids buffer layer regions. Wrinkles extend for several $\mu$m in length and have a height of a few nm and a width of tens of nm (as shown in the inset to Fig.~\ref{Fig3}(a)). They do not follow random directions, but rather the 6-fold symmetry of graphene. Wrinkles of the uppermost graphene sheet are common for epitaxial graphene, yet, they may never appear without intercalation.\cite{NDiaye2009,Sutter2009,buss2023,buss2025} In the present case, however, wrinkles were not observed on the surface prior to Rb evaporation, but readily appeared after Rb evaporation. Thus, here, the wrinkles observed are due to Rb atoms which are likely intercalated below the sample surface and sit between the graphene and the buffer layer (since wrinkles avoid the buffer layer).
In addition to the wrinkles, quite flat areas appear, shown and labeled as RbG in Fig.~\ref{Fig3}(b). These areas are Rb-intercalated graphene regions which prevalently extend from the wrinkles and protrude $\sim325$~pm above the monolayer graphene (as shown in the inset to Fig.~\ref{Fig3}(b)). On the buffer layer, we only found randomly distributed Rb atoms adsorbed on the surface and no indication pointing towards intercalation. Low-temperature Rb deposition results in the same sample morphology, with the additional presence of an extensive disordered Rb adlayer on the monolayer graphene regions. 

\begin{figure}[hbt!]
	\centering
	\includegraphics[width=0.5\linewidth]{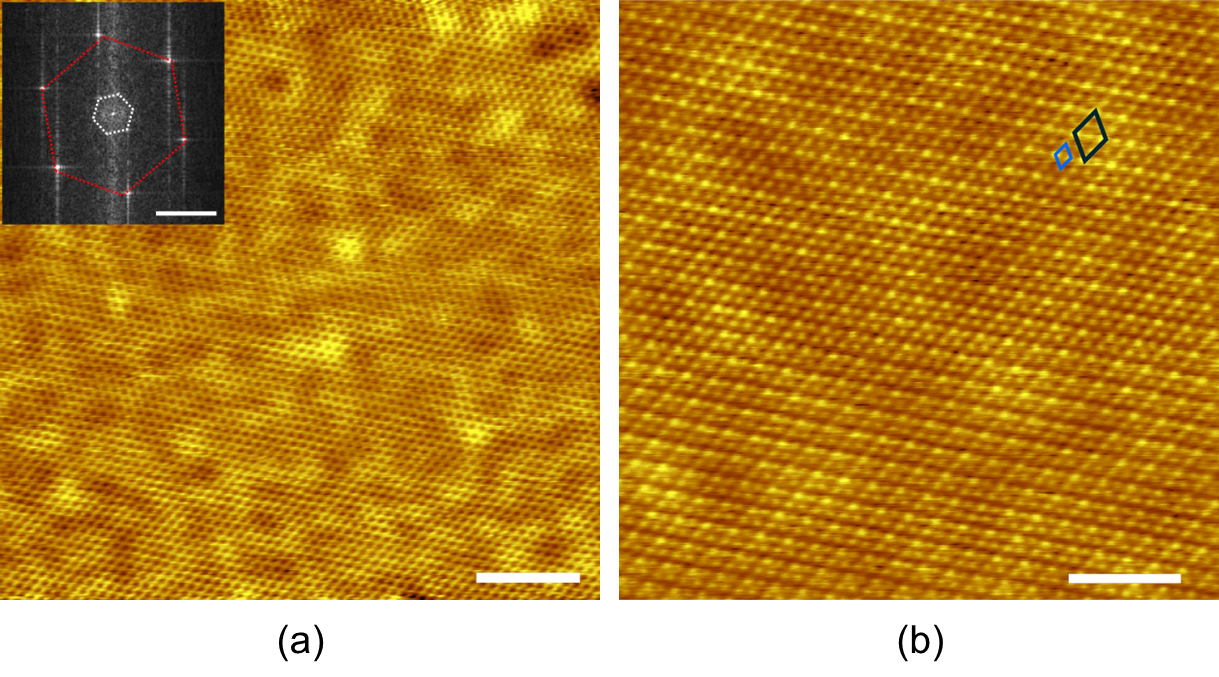}%
	\caption{STM topographic images of Rb-intercalated regions in monolayer graphene showing a Rb$(2\times2)$ ordering. (a) STM scan showing the $6\sqrt{3}$-moirè modulation and the Rb$(2\times2)$ ordering highlighted in the corresponding FFT shown in the inset by white and red hexagons, respectively. (b) Atomically resolved STM scan showing the Rb$(2\times2)$ arrangement together with the graphene lattice, highlighted by black and blue rhombi, respectively. Scan size: (a) $(28\text{~nm}\times28\text{~nm})$ and (b) $(10\text{~nm}\times10\text{~nm})$. Scale bar: (a) 5~nm, inset to (a) 2~nm$^{-1}$, and (b) 2~nm.}
	\label{Fig4}
\end{figure}

\begin{figure}[htpb!]
	\centering
	\includegraphics[width=0.4\linewidth]{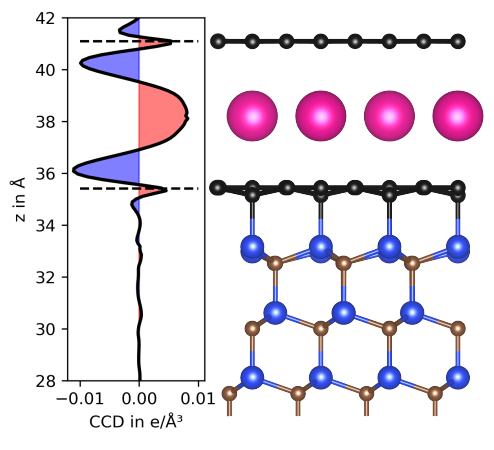}%
	\caption{Charge density difference (CDD) and optimized geometry of the Rb intercalated $(2\times 2)$ structure in monolayer graphene on SiC(0001) obtained from DFT analysis. Due to Rb intercalation, the graphene-buffer layer separation increases to 5.77~\AA, which is in good agreement with results we obtained by STM for the Rb-$(2\times 2)$ reconstruction.}
	\label{Fig5}
\end{figure}

RbG areas often show the morphology reported in Fig.~\ref{Fig4}. These topographies show the presence of a weak long-range $6\sqrt{3}$ modulation, also visible in the fast-Fourier-transform (FFT) of the STM image (inset to Fig.~\ref{Fig4}(a)), which evidences that the bonds between the SiC substrate and the buffer layer are still intact. As indicated in Fig.~\ref{Fig4}(b), the graphene lattice is resolved all over the investigated area, suggesting that Rb atoms do not rSIde on the surface but rather below the topmost graphene layer, consistent with the strong graphene diffraction spots seen in Fig.~\ref{Fig2}. A larger hexagonal pattern showing a peak-to-peak distance of 492~pm can be resolved, as well. This latter is consistent with a $(2 \times 2)$ ordering with respect to the graphene lattice. The local variation in height is at most 30 pm, which is much smaller than the ionic radius of Rb atoms (152~pm),\cite{Lide2004CRC} again compatible with a Rb intercalated phase. The $(2 \times 2)$ RbG regions have an apparent height difference relative to the non-intercalated monolayer graphene of $(301\pm68)$~pm, which is in good agreement with the layer separation we obtained by DFT for a single $(2 \times 2)$-Rb interlayer (as reported in Fig.~\ref{Fig5} and in Section~S5 in the SI). Thus, the $(2 \times 2)$ structure is formed by a single Rb interlayer sandwiched between monolayer graphene and the buffer layer. 

\begin{figure}[hbt!]
	\centering
	\includegraphics[width=0.5\linewidth]{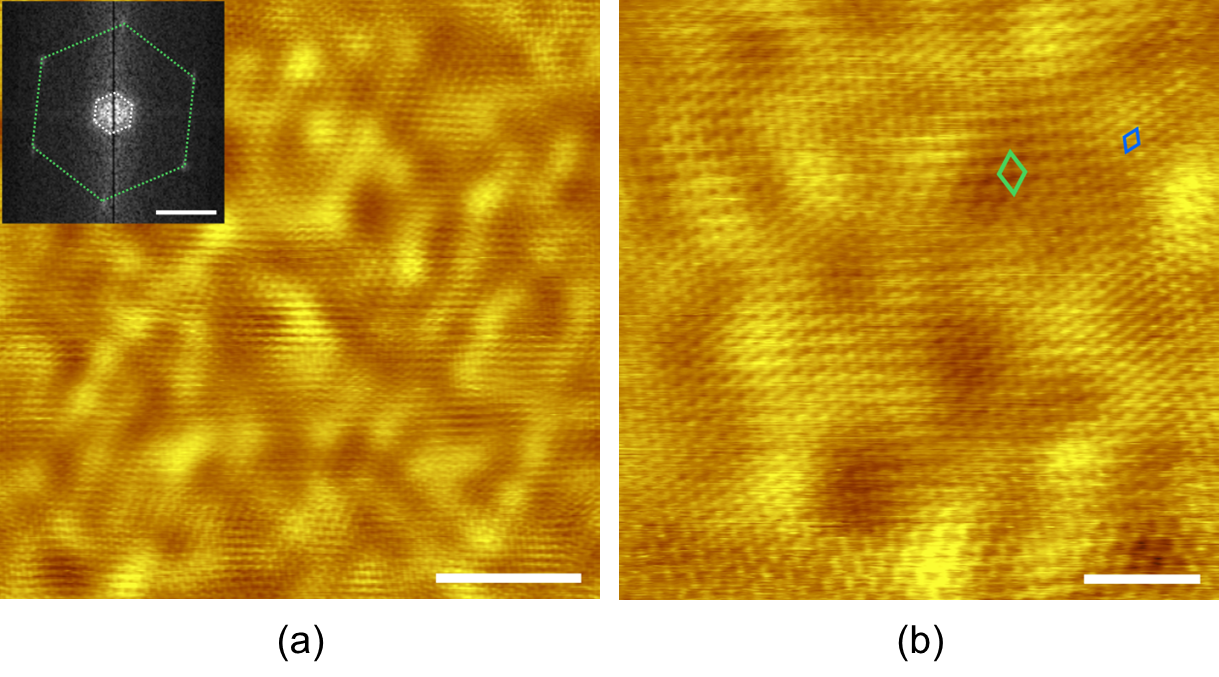}%
	\caption{STM topographic images of Rb-intercalated regions in monolayer graphene showing a Rb$(\sqrt{3}\times \sqrt{3})$R30° ordering. (a) STM scan showing the $6\sqrt{3}$-moirè modulation and the Rb$(\sqrt{3}\times \sqrt{3})$R30° ordering highlighted in the corresponding FFT shown in the inset by white and green hexagons, respectively. (b) Atomically resolved STM scan showing the Rb$(\sqrt{3}\times \sqrt{3})$R30° arrangement together with the graphene lattice, highlighted by green and blue rhombi, respectively. Scan size: (a) $(30\text{~nm}\times30\text{~nm})$ and (b) $(10\text{~nm}\times10\text{~nm})$. Scale bar: (a) 5~nm, inset to (a) 2~nm$^{-1}$, and (b) 2~nm.}
	\label{Fig6}
\end{figure}

Other RbG regions reveal the morphology shown in Fig.~\ref{Fig6}. 
Again, these topographies show the presence of a weak long-range $6\sqrt{3}$ modulation, also visible in the FFT of the STM image (inset to Fig.~\ref{Fig6}(a)). As seen from Fig.~\ref{Fig6}(b), the graphene lattice is observed all over the surface, and a larger hexagonal pattern showing a peak-to-peak distance of 426 pm and a unit cell rotated by 30° with respect to the graphene is observed, as well. This latter structure is consistent with Rb intercalated under the topmost graphene surface with a $(\sqrt{3} \times \sqrt{3})$R30° ordering with respect to the graphene lattice. The $(\sqrt{3} \times \sqrt{3})$R30° RbG regions have an apparent height difference measured with respect to the non-intercalated monolayer graphene of $(354\pm43)$~pm, a number again compatible with only a single alkali-metal intercalated layer. The slightly higher vertical spacing measured for the $(\sqrt{3} \times \sqrt{3})$R30° compared to the $(2 \times 2)$ morphology,  might be a result of the difference in the lateral compression between the two structures. The distance between Rb atoms in the $(2 \times 2)$ structure (492~pm) is close to that of Rb in the bulk (484~pm).\cite{Grosso2013} In the case of the $(\sqrt{3} \times \sqrt{3})$R30°, the distance between Rb atoms is 426~pm, and thus the lattice is laterally compressed by $\sim10\%$ with respect to the bulk form. This would in turn increase the vertical spacing, thus the distance between the buffer layer and monolayer graphene. It results that also the $(\sqrt{3} \times \sqrt{3})$R30° superstructure is due to a single Rb interlayer sandwiched between the monolayer graphene and the buffer layer. 

Fully in agreement with the LEED characterization, both $(2\times2)$ and $(\sqrt{3}\times \sqrt{3})$R30° Rb ordered superstructures are resolved by STM. These correspond to a single ordered 2D-alkali metal interlayer arranged below the topmost graphene surface.  

\begin{figure}[hbt!]
        \centering
        \includegraphics[width=0.5\linewidth]{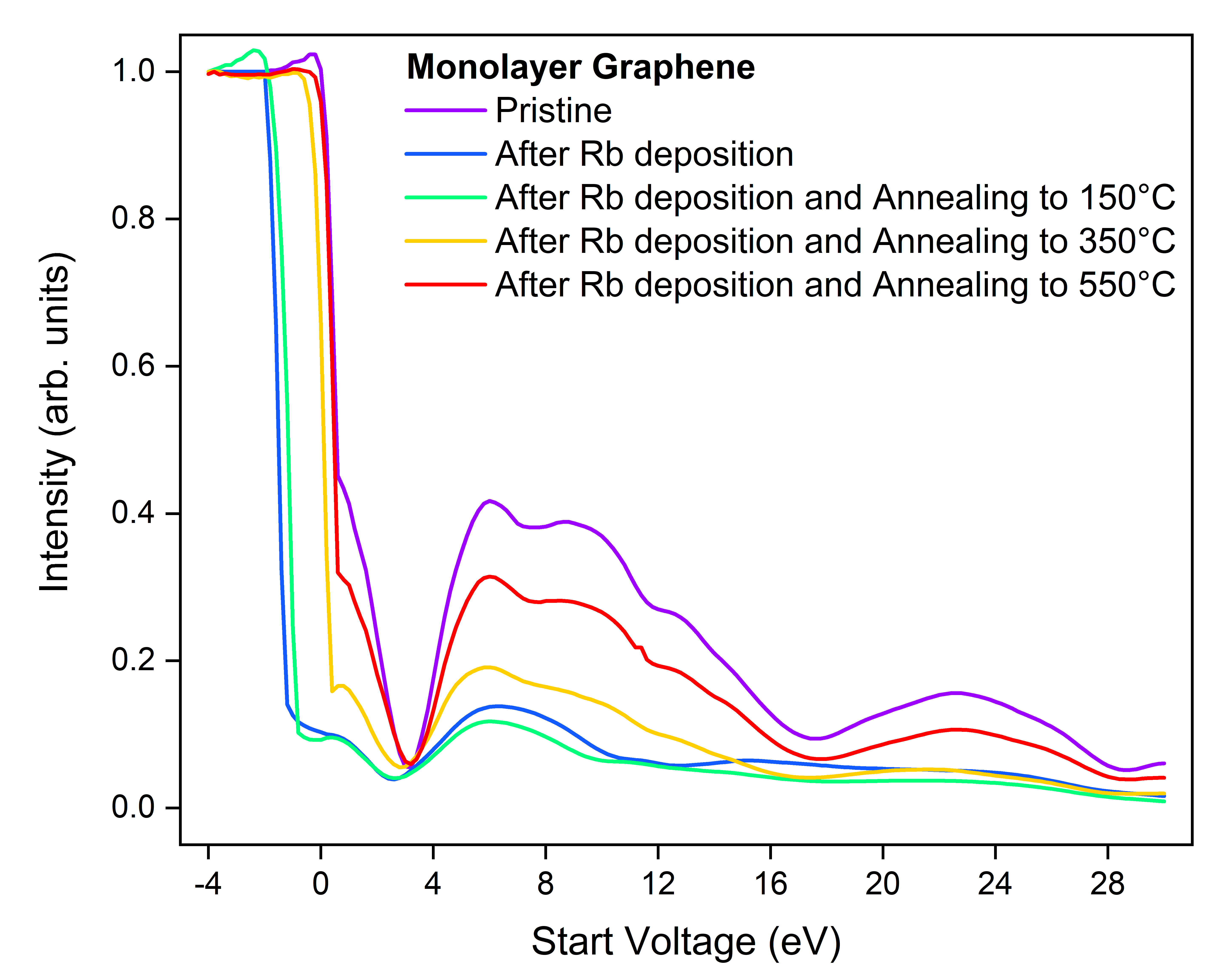}%
        \caption{LEEM-IV spectra of monolayer graphene obtained after Rb deposition at room temperature followed by annealing cycles. The spectra were extracted and averaged over regions of interest of approximate size $(250\text{~nm}\times250\text{~nm})$ and normalized to the intensity of the threshold energy for total reflection of electrons. The LEEM-IV spectrum of pristine monolayer graphene has been included as a reference.}
	\label{Fig7}
\end{figure}

Additional insight comes from LEEM analysis, reported in Fig.~\ref{Fig7} and in Fig.~S5 in the SI. LEEM is a well-suited technique for the investigation of intercalation processes. The energy-dependent coupling of the incoming electrons to the graphene interlayer states leads to a characteristic modulation of the reflectivity, where the number of dips in the intensity-voltage (LEEM-IV) spectrum corresponds to the number of freestanding graphene layers.\cite{hibino2008} Therefore in strongly bound graphene systems, successful intercalation and decoupling of the graphene layer increases the number of dips by one.\cite{Sutter2013, Coletti2011} Before Rb deposition, the surface consists mainly of monolayer graphene and buffer layer, with the monolayer showing the characteristic reflectivity dip in the LEEM-IV around 4~eV. After Rb deposition, the LEEM-IV spectra are attenuated, but the number of dips is unchanged in both the monolayer and the buffer layer regions. Therefore, we can exclude intercalation of Rb at the buffer layer/SiC interface. Additionally, a shift in the position of the dip by $\sim 0.5$~eV towards lower energies in the LEEM-IV spectrum is observed. As the energetic position of the graphene interlayer states strongly depends on the layer distance~\cite{feenstra2013,srivastava2013} such a shift might be related to a change in the interlayer spacing of the graphene due to Rb intercalation. However, such an interpretation of the LEEM-IV spectra is difficult as it does not take into account the influence of the Rb interlayer on the electronic properties of the system, which affect the reflectivity. Nevertheless, LEEM further supports the LEED and STM conclusion that Rb is intercalated and rSIdes between monolayer graphene and the buffer layer, without intercalating the buffer layer.

Furthermore, after Rb intercalation, the threshold energy for total reflection of electrons from the surface is shifted by approximately 2~eV to lower energies, indicating a change in the work function by the same amount. This shift is larger than the work function change of 0.5~eV calculated by DFT for Rb intercalation (see Section~S6 in the SI) and more comparable to the work function change expected for Rb adsorption, but based on our STM results we can rule out adsorption of Rb on monolayer graphene. At the threshold voltage, the kinetic energy of the incoming electrons is minimal directly above the surface. This makes them susceptible to lateral electric fields emerging between regions with different work functions on the surface, which leads to overestimation of work function shifts in samples with laterally varying work functions.\cite{jobst2019} As STM shows that the graphene monolayer is not homogeneously intercalated, it is likely that the lateral fields between intercalated and non-intercalated regions lead to the larger apparent work function shift. Nevertheless, the negative shift of the apparent work function still confirms substantial n-type charge transfer doping of graphene by Rb intercalation, which, from DFT results, we estimate to result in a downward shift of the Dirac point by approximately 1.1~eV.

Selected annealing steps above RT provide information about Rb diffusion and de-intercalation kinetics.  
As graphene samples dosed with Rb at RT are heated just above room-temperature (50~°C), any sign of Rb ordering ($(2 \times 2)$ or $(\sqrt{3}\times \sqrt{3})$R30°) disappears from the diffraction pattern. This is also confirmed by STM imaging which reveals that Rb is still intercalated as a single layer underneath the topmost graphene surface and above the buffer layer, but without apparent order (as reported and further discussed in Section~S4 in the SI). 

\begin{figure}[hbt!]
	\centering
	\includegraphics[width=0.5\linewidth]{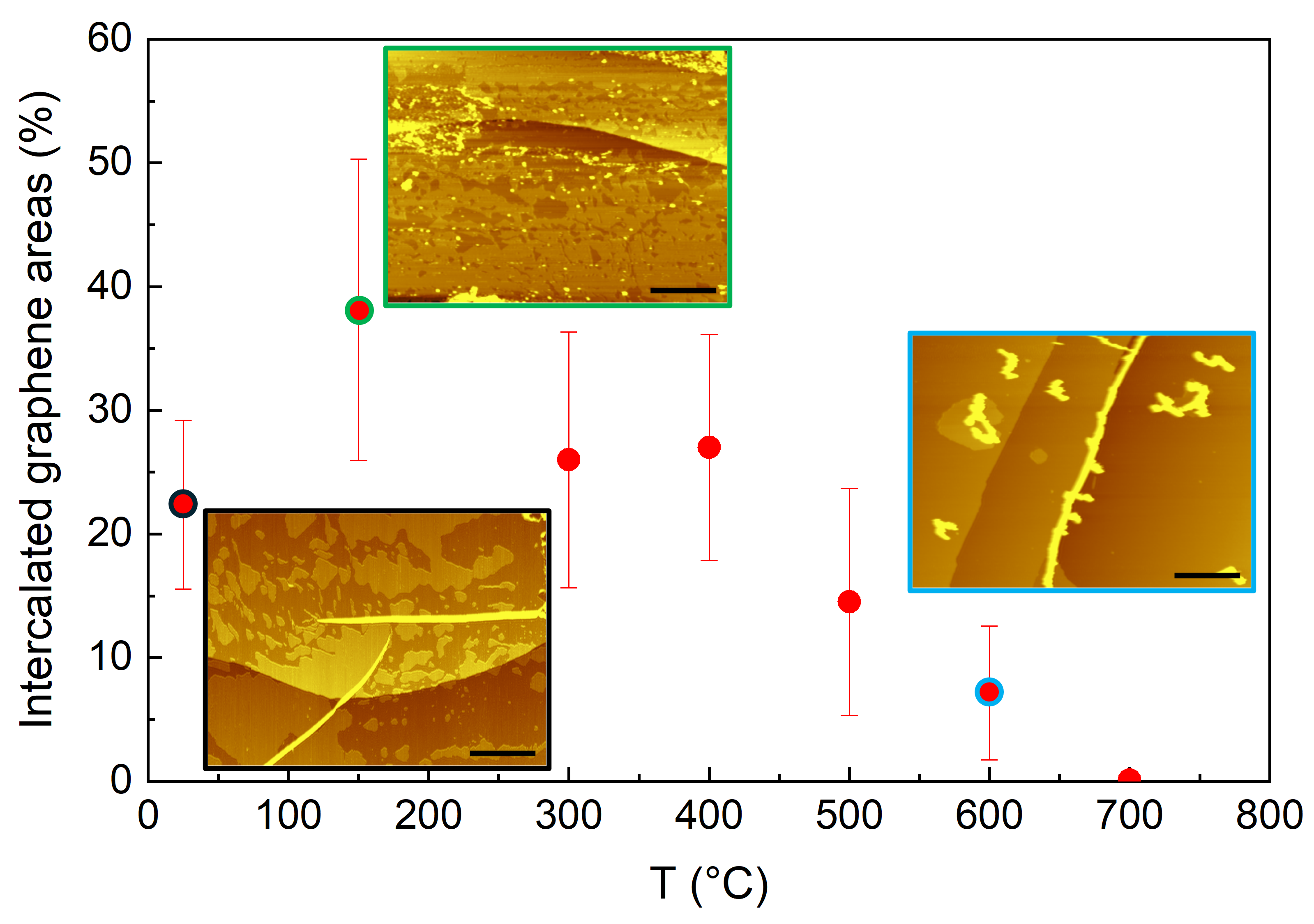}%
	\caption{Statistical analysis of several STM images $(500\text{~nm}times\text{500~nm})$ reporting the fraction of Rb intercalated graphene area as a function of the annealing temperature (T). Inset to the graph: STM images of size $(500\text{~nm}\times380\text{~nm})$ representative of the sample surface after Rb deposition at RT (black), and subsequent annealing up to 150~°C (green) and 600~°C (blue). Colors for the frames of STM images have been used following to the colors of the data points highlighted in the plot.}
	\label{Fig8}
\end{figure}

Analysis of the Rb-intercalated graphene area obtained from large-scale STM imaging, reported in Fig.~\ref{Fig8} (and Section~S4 in the SI), shows the presence of two temperature regimes. 
At first, we observe an increase in the amount of intercalated areas. Indeed, after annealing the sample at 150~°C, the intercalated area almost doubles with respect to the initial fraction (from $\sim20$\% to $\sim40$\%). This requires a supply of Rb, i.e., for this to happen, there must be a Rb source in the sample. Since no adsorbed Rb atoms or clusters are observed after deposition at room temperature, Rb is likely provided by the wrinkles. Still, diffusion of the ordered intercalated Rb atoms may happen, also leading to an enlargement of the intercalated region.

At temperatures above 150~°C, desorption sets in. The fraction of the intercalated area decreases, and the single intercalated islands shrink in size and fragment. Rb clusters, not present before, start appearing on the graphene surface. Already at 300~°C the area fraction of the intercalated islands returns to the RT value. The desorption process continues up to approximately 700~°C, when the intercalated fraction reduces to zero. The surface is left with wrinkles, a few dispersed clusters, and sparse Rb multilayered islands. This demonstrates that the intercalation process is reversible.
The LEEM analysis shown in Fig.~\ref{Fig7} reveals consistent results and confirms the reversibility of the intercalation process on a different length scale. With annealing cycles above 350~°C the shape of the LEEM-IV progressively changes towards that of pristine graphene. After annealing the sample at 550~°C, the work function as well as the position of the dip return to the values of the pristine graphene, indicating Rb desorption and retrieval of the initial pristine surface.

\section{Discussion}

Changes in the diffraction pattern revealed a phase transition from a well-ordered Rb $(2 \times 2)$ structure to a more closely packed Rb $(\sqrt{3} \times \sqrt{3})$R30° structure. Both structures, as confirmed by STM, LEED and LEEM, correspond to the formation of a well-ordered alkali-metal layer intercalated below the topmost graphene layer, i.e., between the buffer layer and monolayer graphene. The phase transition is obtained by increasing the Rb coverage (shown in the case of RT-Rb deposition) or by allowing diffusion to take place with annealing cycles (shown in the case of LT-Rb deposition). At the same time, we observe no evidence of Rb intercalation below the buffer layer, but only disordered adsorption of Rb atoms on the surface of the buffer layer. On the other hand, on monolayer graphene, we only observe intercalation of Rb atoms and no adsorption on the monolayer surface at RT. This observation aligns with our results from DFT, which show that intercalation of Rb atoms between the buffer layer and monolayer graphene is an energetically favorable process compared to adsorption on the monolayer graphene surface. At the same time, DFT shows that adsorption of Rb atoms on the buffer layer surface is more favorable than Rb intercalation at the buffer layer/substrate interface.

The $(2 \times 2)$ structure is analogous to that of bulk C$_8$Rb. Previous experiments on Rb deposition onto graphite showed that Rb atoms readily intercalate the graphite surface above 80~K, forming a $(2 \times 2)$ monolayer under the topmost graphene layer before further diffusing into the bulk.\cite{Igarashi2023, Caragiu2005, Algdal2007} Also experiments on epitaxial bilayer graphene grown on SiC(0001) reported that Rb atoms readily intercalate at 80~K between the two graphene layers with a $(2 \times 2)$ ordering.\cite{Kleeman2013, Kleeman2014} However, for Rb atoms deposited on epitaxial monolayer graphene (i.e., the same system investigated in this work), either at room-temperature~\cite{Watcharinyanon2011} or at low temperature~\cite{Watcharinyanon2011, Kleeman2013}, no superstructure was observed previously, which is in apparent contrast to our results. 
However, we have indeed shown that the $(2 \times 2)$ diffraction pattern obtained by room-temperature Rb deposition is rather unstable, and completely disappears within 20~min. This lack of stability may suggest why this phase was not observed in previous reports. In turn, when Rb is deposited at low-temperature, the $(2 \times 2)$ diffraction pattern appears with low intensity and with a high background. In addition to the presence of an ordered Rb intercalated structure, there is an amorphous Rb overlayer that fully covers the sample surface. This latter might have, again, hindered the visualization of ordered phases in previous reports.

The emergence of a $(\sqrt{3} \times \sqrt{3})$R30° periodicity produced by a single intercalated Rb layer is quite unexpected and a novelty. The $(\sqrt{3} \times \sqrt{3})$R30° structure pertains to a Rb-Rb distance of 426~pm which corresponds to compression of the Rb lattice by $\sim10\%$ compared to the nearest neighbor distance of Rb atoms in their bulk crystal structure (484~pm).~\cite{Grosso2013} However, such an intralayer compression is feasible when the Rb intercalated atoms are almost completely ionized, with the electrons rSIding in the neighboring graphene and buffer layer.~\cite{Kaneko2017,Kleeman2013} The $(\sqrt{3} \times \sqrt{3})$R30° structure has been frequently observed for smaller AMs (e.g., Li and Na) intercalated in the C$_6$AM form both in graphite~\cite{Caragiu2005} and epitaxial monolayer graphene~\cite{Fiori2017}. In the latter system, the Li-$(\sqrt{3} \times \sqrt{3})$R30° structure forms between between the graphene and the buffer layer, after the buffer layer is intercalated with Li and detached from the SiC. Recently, it has been shown that the formation of AM bilayers between two graphene sheets, corresponding to a C$_6$AM$_2$C$_6$ configuration, also leads to $(\sqrt{3} \times \sqrt{3})$R30° periodicity.~\cite{Lin2024} However, we can exclude such a mechanism for Rb-intercalated monolayer graphene/SiC, as the resulting layer separation would be too large to be compatible with the layer separation that we measure via STM. 

Now, based on these experimental observations, we can understand how Rb intercalation in monolayer graphene occurs. Rb has sufficient mobility on the graphene surface, even at temperatures as low as 100~K, so that it readily forms a wrinkles network. This represents the first intercalation stage of Rb under the topmost graphene layer and likely occurs at the SiC steps and graphene defects.
When Rb is deposited at low temperatures, the sticking coefficient of the surface is higher than at RT. A fraction of Rb atoms readily intercalate in extended areas and arrange with a $(2 \times 2)$ periodicity, and the remaining Rb atoms are amorphously adsorbed on the surface. Since diffusion is limited at low temperature, the structure is stable. As the sample temperature is increased towards room temperature, Rb atoms can more easily diffuse, and those in the amorphous \textit{over}layer can intercalate as well. This leads to a change in the structural arrangement of Rb atoms in the intercalated areas which become more densely packed and show a $(\sqrt{3} \times \sqrt{3})$R30° ordering. 

On the other hand, when Rb is evaporated at room temperature, there is a balance between adsorption and desorption of Rb atoms. All Rb atoms that stick to the surface are readily intercalated. The higher diffusion rate of Rb atoms at RT establishes a dynamic equilibrium of the intercalated atoms beneath the graphene surface. This dynamic equilibrium gives rise to a metastable $(2 \times 2)$ ordering. However, once the density of neighboring Rb atoms is high enough upon deposition of further material, intercalated Rb atoms arrange into a stable closely packed $(\sqrt{3} \times \sqrt{3})$R30° structure. 

All these processes occur between graphene and the buffer layer. Intercalation of Rb atoms below the buffer layer would require breaking the (partial) covalent bonding between the buffer layer and the SiC substrate. Most metals acquire the required energy to decouple the buffer layer via high temperature annealing.\cite{Briggs2019} However, in the case of Rb we have shown that de-intercalation and desorption already start at temperatures above 150~°C. Below this temperature, Rb atoms may not have enough energy to diffuse below the buffer layer and decouple it from the substrate. In addition, our results from DFT calculations (see Section~S7 in the SI) suggest that intercalation of the buffer layer with Rb is energetically unfavorable due to the size of the Rb atoms. Since the buffer layer is (partially) covalently bound to the SiC, it is necessary for the intercalant to saturate the resulting dangling bonds of the SiC after decoupling from the substrate, as in the case of intercalation of the buffer layer with Li.\cite{Fiori2017} However, the large nearest neighbor distance of Rb does not allow the saturation of all dangling bonds, making the intercalation of the buffer layer energetically unfavorable.

\section{Conclusions}

We have showcased a coverage and temperature-dependent experiment of Rb intercalation under epitaxial monolayer graphene on SiC(0001). Using LEED, STM, LEEM, and $\mu$-LEED measurements supported by DFT calculations, we have demonstrated that Rb atoms intercalate the topmost graphene layer, but not the buffer layer. The intercalated Rb atoms form an alkali metal interlayer which shows two different periodicities compared to the graphene lattice, i.e., a $(2 \times 2)$ and a $(\sqrt{3} \times \sqrt{3})$R30° structure. Rb intercalation almost doubles the distance between monolayer graphene and the buffer layer, as confirmed by DFT analysis, and induces an extensive doping of the graphene.
By performing annealing cycles at high temperatures, we have shown that after a first stage in which diffusion prevails over desorption and the intercalated areas expand, desorption sets in, and at approximately 600~°C we could retrieve the original surface, demonstrating the reversibility of the intercalation process.

\section*{Author contributions}
L. F. - Conceptualization, Investigation, Formal analysis, Visualization, Writing - Original Draft, Writing - Review \& Editing; S. V. - Conceptualization, Investigation, Validation, Writing - Review \& Editing; T. O. M. - Investigation, Resources, Writing - Review \& Editing; L.B.: Investigation, Formal analysis, Writing - Review \& Editing; A. R., N. M., and C. C.: Resources, Writing - Review \& Editing; J. I. F. - Validation, Writing - Review \& Editing; A. L. - Investigation, Supervision, Resources, Writing - Review \& Editing; S. H. - Conceptualization, Resources, Validation, Writing - Review \& Editing, Supervision, Project administration.

\section*{Conflicts of interest}
There are no conflicts to declare.

\section*{Data availability}

The data supporting this article have been included as part of the Supplementary Information.

\section*{Acknowledgements}

We acknowledge financial support from the PNRR MUR Project PE0000023-NQSTI founded by the European Union-NextGenerationEU.

\end{document}